\documentclass[journal]{IEEEtran}
\usepackage{graphicx}
\usepackage[cmex10]{amsmath}

\usepackage{framed}
\usepackage{cite}

\begin{document}
\title{The Sample Allocation Problem and Non-Uniform Compressive Sampling}
\author{Andriyan B.~Suksmono%
\thanks{A.B. Suksmono is with the School of Electrical Engineering and Informatics, Institut Teknologi Bandung, Jl.Ganesha No.10, Bandung, Indonesia, e-mail:suksmono@stei.itb.ac.id.}
\thanks{This paper has been partially presented in ... }
\thanks{Manuscript received ... ; revised ... }}

\markboth{A.B.Suksmono: The S.A.P. and Non Uniform C.S.} %
{Shell \MakeLowercase{\textit{et al.}}: The Sample Allocation Problem and Non-Uniform Compressive Sampling}

\maketitle

\begin{abstract}
This paper discusses sample allocation problem (SAP) in frequency-domain Compressive Sampling (CS) of time-domain signals. An analysis that is relied on two fundamental CS principles; the Uniform Random Sampling (URS) and the Uncertainty Principle (UP), is presented. We show that CS on a single- and multi-band signals performs better if the URS is done only within the band and suppress the out-band parts, compared to ordinary URS that ignore the band limits. It means that sampling should only be done at the signal support, while the non-support should be masked and suppressed in the reconstruction process. We also show that for an $N$-length discrete time signal with $K$-number of frequency components (Fourier coefficients), given the knowledge of the spectrum, URS leads to exact sampling on the location of the $K$-spectral peaks. These results are used to formulate a sampling scheme when the boundaries of the bands are not sharply distinguishable, such as in a triangular- or a stacked-band- spectral signals. When analyzing these cases, CS will face a paradox; in which narrowing the band leads to a more number of required samples, whereas widening it leads to lessen the number. Accordingly; instead of signal analysis by dividing the signal’s spectrum vertically into bands of frequencies, slicing horizontally magnitude-wise yields less number of required sample and better reconstruction results. Moreover, it enables sample reuse that reduces the sample number even further. The horizontal slicing and sample reuse methods imply non-uniform random sampling, where larger-magnitude part of the spectrum should be allocated more sample than the lower ones.
\end{abstract}

\begin{IEEEkeywords}
compressive sampling, CS-SFCW radar, Non Uniform Sampling, ultrawideband radar.
\end{IEEEkeywords}

\section{Introduction}

\IEEEPARstart{C}{ompressive} sampling (CS) deals with partial measurement on a signal and guarantees that if a minimum number of required samples are given, although it is much fewer than normally required in conventional sampling, a sparse signal can be exactly reconstructed with overwhelming probability \cite{Donoho2006}, \cite{CandesTao2006}, \cite{CandesRombergTao2006}. The CS measurements are conducted in a particular signal domain. In some sensing devices, such as MRI (Magnetic Resonance Imaging), Radio-Interferometric Telescope, or SFCW (Stepped Frequency Continuous Wave Radar), measurements are conducted in frequency domain and yields a number of Fourier coefficients. The CS version of these devices selects the available coefficients in a uniform random fashion. When we know that the signal occupies only a particular part of the frequency, should we do the sampling to the whole range? Furthermore, if the typical spectrum of the signal is known, how this knowledge guide us to select the best set of samples, in term of less sample number for a comparable reconstruction result ?

In the conventional transform-based signal compressions, higher-energy coefficients are more prioritized than the lower ones. In the DCT (Discrete Cosine Transform) coding, for examples, high energy coefficients are allocated more bits than the lower ones. Moreover, the bit allocation scheme can even be standardized since most of signals under interest have similar frequency characteristic; that is to say, the signals have a typical spectrum. In contrast, CS does not introduce such preferences. The (so called) democratic principle of CS dictates that the sample selection is done in a non-adaptive fashion, which means that uniform random sampling (URS) is applied to select the samples and all of them are equally important.  

Adaptive sampling is actually found to be useful in some practical applications. In \cite{Suksmono2009a} we have shown that EES (Equipartition of Energy Sampling) scheme can reduce sample number while maintaining comparable reconstruction quality. This technique may significantly improve the acquisition speed of CS-based SFCW (Stepped-Frequency Continuous Wave) radars \cite{Suksmono2008a}. The EES is also adopted by \cite{MiaoEtAl2013} to accelerate the measurement process in MRI, while also maintaining similar quality. Similar cases will also be found in CS-based radio-interferometric imaging telescopes \cite{Suksmono2009b}, \cite{WiauxEtAl2009}, where the selection of best sites or configuration of VLBI (Very Long Baseline Interferometry) stations is a crucial issue. 

The EES scheme addresses the sample allocation problem (SAP) issue by proposing that the density of the sampling points is proportional to the density of the Fourier magnitude coefficient (spectrum). The idea has been derived heuristically by an analogy with the conventional transform-coding principle that high energy coefficients are more important than the lower ones. How can such a proposition be justified by and compatible with the CS principles? A better explanation that is consistent with the CS principles, instead of adopting classical considerations, will be crucial for both of practical and theoretical development of the CS.

The SAP problem can be stated as follows: \textit{given a particular prior knowledge of a signal and a small number of samples, what is the best set of sampling points that maximize the reconstruction performance?} In this paper, the prior knowledge to be addressed is (typical) magnitude spectrum or distribution of Fourier coefficients of the signal, whereas the performance is measured by RMSE (Root Mean Squared Error). In particular, we consider temporal/spatial signal that are sampled in frequency domain. The objective is to reconstruct the signal from a given set of the Fourier coefficients. 

The rest of the paper are organized as follows. In Section \ref{Section2}, we discuss frequency domain sampling for various kinds of signals: single- and multiple- time domain impulse, single- and multiple- tone signals, and single- and multiple- band signals. We analyze and verify how the URS works for each of these cases, and how to modify the projection operator in sampling and reconstruction. In Section \ref{Section3}, the results from the previous Section are used to further analyze when the band limit is not sharp, such as in triangular- and stacked multi-band-spectral signals. The proposed technique is simulated and analyzed in Section \ref{Section4}. Then, we conclude the paper in Section \ref{Section5}.

\section{Frequency-Domain Compressive Sampling}
\label{Section2}

\subsection{Operations on the Identity Matrix}
\label{SubSection2A}
In the CS, the sampling process is mathematically expressed as projection of the signal into a particular bases, which is performed by a projection or sensing matrix $\Phi$. Since we only need to perform random selection of available frequency components, we use the simplest zero-one matrix to do the task. We will find that $\Phi$ is actually obtained from the identity matrix by a few operation(s) introduced in this sub section.

\subsubsection{Identity matrix $I$ and delta functions $\left\{\delta_k \right\}$}
Throughout the paper, we will be working with an $N$-length discrete time signal. We will use $\delta_k$ to denote a unit impulse (Kronecker’s delta) that is located at time index $k$; therefore, it is a column vector of length $N$ with zero entries except at the $k^{th}$ row whose values is unity, i.e.,
\begin{equation}
  \label{Eq1}
  \delta_k^T = \bordermatrix{~ & 1 & 2 &\cdots &k-1 &k &k+1 & \cdots &N \cr
                  & 0 & 0 &... &0 &1 &0 &... &0 
         		}
\end{equation}
The $(...)^T$ denotes transpose operation. Using this notation, an $N \times N$ identity matrix can be expressed as
\begin{equation}
  \label{Eq2a}
  I = \begin{bmatrix}
  		1 &0 &\cdots &0 \\
		0 &1 &\cdots &0 \\
		\cdots &\cdots &\cdots &\cdots \\
		0 &0 &\cdots &1 \\
  \end{bmatrix}
   = \begin{bmatrix}
   \delta_1 &\delta_2 &\cdots \delta_N
   \end{bmatrix} 
\end{equation}
and equivalently, it can also be expressed as 
\begin{equation}
  \label{Eq2b}
  I = \begin{bmatrix}
  	 \delta_1 &\delta_2 &\cdots \delta_N
   \end{bmatrix}^T 
\end{equation}

\subsubsection{Deletion of $I$ by $\left\{\delta_k\right\}$ }
Deletion of the $k^{th}$ column of the identity matrix by a (single) delta function $\delta_k$ that produce an $N \times (N-1)$ zero-one matrix is an operation that is performed as follows
\begin{equation}
  \label{Eq3a}
  	I -\delta_k = \left[ \underbrace{ \delta_1 \; \delta_2 \;\cdots \; \delta_{k-1} \; \delta_{k+1} \; \cdots \; \delta_N }_\text{N-1} \right]
\end{equation}
This result is also denoted as 
 \begin{equation}
 \label{Eq3b}
   \delta_k^C \equiv I-\delta_k
 \end{equation}
Similarly, deletion of the identity matrix by a set of $K$ number of distinct delta functions $D=\left\{ \delta_{i_1}, \cdots, \delta_{i_K} \right\}$ , where $i_1 \neq i_2 \neq \cdots \neq i_K$, produces an $N \times (N-K)$ matrix $D^C$  as

\begin{equation}
  \label{Eq4}
  I-D=I-\delta_{i_1}-\delta_{i_2}-\cdots-\delta_{i_K} = D^C
\end{equation}

\subsubsection{Extension of $\left\{\delta_k \right\}$ and vectors}
An extension of the set of unit impulse $D$, denoted by $\bar{D}$, is a rearrangement of the identity matrix, in which the first $K$-rows is $D^T$ and the remaining ones are $\left(I-D\right)^T = \left(D^C\right)^T$ , i.e., it is an $N \times N$ matrix constructed as 
\begin{equation}
  \label{Eq5}
 \bar{D} = \begin{bmatrix}
 	D^T \\
 	\left(D^C \right)^T
 	\end{bmatrix}
 	= \left[ \underbrace{\delta_{i_1} \; \cdots \; \delta_{i+k}}_{K}  \underbrace{\; \delta_1 \; \delta_2 \; \cdots \; \delta_{K}}_{N-K} \right]
\end{equation}
This definition implies that the extension of $\delta_1$ yields the identity matrix, i.e., $\bar{\delta_1}=I$ .

We also define an extension of a vector $v=\left[ v_1 \; v_2 \; \cdots \; v_K \right]^T$of length $K<N$, denoted by $\bar{v}$, as appending $N-K$ number of zeros to the vector, so that the length becomes $N$, i.e.,
\begin{equation}
  \label{Eq6}
  v=\left[ \underbrace{ v_1 \; v_2 \; \cdots \; v_k \; 0 \; 0 \; \cdots \; 0}_{N} \right]
\end{equation}

\subsection{Fourier Transform Pair of Discrete Time Signal and the Basic CS Theory}

We will be using the time index $n$ and frequency index $k$ explicitly in the signal, such as $s(n)$ or $S(k)$, when required, or drop them to become $s$ or $S$ as an $N$-length vector. Let $s(n)$ be an $N$-length time-domain signal whose discrete Fourier transform (DFT) pair is $S(k)$. The complete set of the DFT bases can be arranged as column vectors of an $N \times N$ DFT matrix $\Psi$,

\begin{equation}
\label{Eq7x}
  \Psi = \frac{1}{\sqrt{N}} 
  \begin{bmatrix}
  	1	&1	&\cdots 	&1 \\
  	1	&e^{j\frac{2\pi}{N}}	&\cdots  &e^{j\frac{2\pi (N-1)}{N}} \\
  	\cdots  &\cdots  &\cdots &\cdots \\
  	1	&e^{j\frac{2\pi(N-1)}{N}}	&\cdots  &e^{j\frac{2\pi (N-1)(N-1)}{N}}
  \end{bmatrix}
\end{equation}

The DFT pairs of a signal are given by
\begin{equation}
 \label{Eq7a}
	s(n) = \frac{1}{\sqrt{N}} \sum_{k=0}^{N-1} S(k) e^{j \frac{2\pi k}{N}n}
\end{equation}
\begin{equation}
\label{Eq7b}
	S(k) = \frac{1}{\sqrt{N}} \sum_{k=0}^{N-1} s(n) e^{-j \frac{2\pi n}{N}k}
\end{equation}
where $j=\sqrt{-1}$. Simplified further, the transform of the signal given by (\ref{Eq7b}) is written as
\begin{equation}
  \label{Eq8}
    S=\Psi s
\end{equation}

Generally, the problem of signal reconstruction from a given number of subsamples can be formulated into the following two questions (adopted from \cite{Tao2007}):
\begin{itemize}
\item Is it possible to reconstruct $s(n)$ uniquely from $M$-number of frequency domain samples?
\item If so, what is the practical algorithm to find it?
\end{itemize}

In the CS, the signal is observed by taking $M$ out-of $N$ samples by using a projection operator that is represented by an $M \times N$ zero-one matrix $\Phi_\Pi$, which mathematically is expressed as
\begin{equation}
  \label{Eq9}
 S_\Pi=\Phi_\Pi S = \Phi_\Pi \Psi s
\end{equation}
Therefore, $S_\Pi$ is an $M$-length signal. The zero-one entries of the projection matrix are arranged so that $S_\Pi$ contains all of the observed coefficients. According to the CS theory \cite{Donoho2006}, \cite{CandesTao2006}, \cite{CandesRombergTao2006}, reconstruction is ideally performed by selecting a solution with least number of coefficients expressed as
\begin{equation}
  \label{Eq10a}
  \left| s \right|_{l_0} = \sum_{n=0}^{N-1} \left| s_n \right|^0
\end{equation}
by assuming that the signal is sparse; i.e., 
\begin{equation}
  \label{Eq10b}
   \min \left| s \right|_{l_0} subject \; to \; \; \; \Phi_\Pi \Psi s =S_\Pi
\end{equation}

Since there is no practical algorithm that can find the solution of (\ref{Eq10b}), it is replaced by selection of a solution whose sum of magnitude is minimum
\begin{equation}
  \label{Eq11a}
  \left| s \right|_{l_1} = \sum_{n=0}^{N-1} \left| s_n \right|^1
\end{equation}
or any other sparsity criteria, such as minimum total variation (TV). The optimization is conducted by
\begin{equation}
  \label{Eq11b}
   \min \left| s \right|_{l_1} subject \; to \; \; \; \Phi_\Pi \Psi s =S_\Pi
\end{equation}

In the followings, we will discuss various signal reconstruction problems from a few number of known Fourier coefficients. Since both of sampling and reconstruction is involved, we will observe how sampling points are allocated and how we should modify the reconstruction expression given by $L_1$ in (\ref{Eq11b}).

\subsection{CS of A Single- and A Multiple Time-Domain Impulse}
\subsubsection{A Single Time-Domain Impulse}
A single impulse will have a uniform spectrum spread over the frequency domain. The signal reconstruction by IDFT requires all of $N$-sampling points for an exact solution. The CS suggests taking only $M \ll N$ point of sample at random, and then reconstruct by using $L_1$ optimization. In contrast to the single- and multi- tone cases to be addressed later, there will be no preference in the selection of such sampling points. It is quite reasonable to take the frequency at uniform-random manner (URS-Uniform Random Sampling) because the magnitude is also uniform over the frequency.

The projection matrix  $\Phi_P$ is constructed by randomly selects $M$-number of rows in the $N \times N$ identity matrix. The subsamples will be an $M$-length vector
\begin{equation}
  \label{Eq12a}
    S_P = \Phi_P \Psi s
\end{equation}
and the reconstruction can performed by the standard CS as follows
\begin{equation}
  \label{Eq12b}
   \min \left| s \right|_{l_1} subject \; to \; \; \; \Phi_P \Psi s =S_P
\end{equation}
 
\subsubsection{A Multiple-Impulse Signal }

The sampling and reconstruction is done almost similar to the single impulse case, except that the number of sample $M$ is increased. This is a consequence of increasing the degree-of-freedom $K$, referring to the CS formula of required number of sample \cite{CandesRomberg2005}
\begin{equation}
  \label{Eq13}
  M \geq C K \log N
\end{equation}
Although the magnitude distribution is generally not uniform anymore, the process of taking sample is actually also done in a uniform-random manner. If sufficient number of $M$ is achieved, $L_1$ will reconstruct the signal exactly.

\subsection{CS of Band Limited Signals}
In the frequency domain, a band limited signal consists of subsequent non-zero coefficients within a range of frequency. In the sampling stage, CS takes the sampling points randomly over the frequency. When the spectrum is known, the limits of in-band and out-band coefficient are also known. It is not necessary to take sample at the out-band area, since we know all of them are zero. Instead, we do the followings
\begin{itemize}
  \item \textbf{In-band sampling}: perform random sampling only inside the band
  \item \textbf{Out-band suppression}: elaborate the knowledge of out-band’s zero coefficients in the reconstruction
\end{itemize}
Consider a signal $s(n)$ whose Fourier coefficients $S(k)$ has a uniform magnitudes within a particular band $\Omega$,
\begin{equation}
  \label{Eq14}
  \Omega = \left\{ k_0 \; , k_0+1 \; ,\cdots , \; k_0+W-1\right\}
\end{equation} 
where $k_0$ is the starting index of the band and $W=\left| \Omega \right|$ denotes the size of the band, i.e., the bandwidth. This uniformity of the magnitude makes the band limits between the in-band and the out-band coefficients sharply distinguishable. 

For illustration, we will use a simple prototypical signal with flat-frequency spectrum over the bandwidth and localized in the time domain. Assuming that $W$ is an odd number, i.e. $W=2L-1$ with $L>0$, the Fourier transform pair of the signals are,
\begin{equation}
  \label{Eq15a}
  S(k) = \begin{cases}
  			1, k \in \Omega	\\
  			0, otherwise
 	      \end{cases}
\end{equation}
\begin{equation}
  \label{Eq15b}
  s(n) = e^{ j \frac{2\pi}{N}  n(L-\frac{1}{2})} 
                \frac{\sin \left( 2\pi L \frac{n}{N}\right)}{\sin \pi \frac{n}{N}}
\end{equation}
Using this signal, CS is done by sampling in frequency domain and performing reconstruction by $L_1$ minimization on $\left| s \right|$.

\subsubsection{Single-Band Signal}
Reconstruction by IDFT requires all of $W$-number of coefficients in (\ref{Eq14}) to get the exact result. When the bandwidth is sufficiently large, i.e., $W$ is close to $N$, the UP implies that the signal will be sparse in time domain. Then, the sampling problem can be treated like one in time-domain impulse; i.e., we can take much fewer number of sample $M \ll W<N$. In-band sampling will be performed by a projection matrix. By denoting the corresponding locations of selected samples as $\left\{ i_1, \; i_2, \cdots, \; i_M\right\}$, the projection matrix $\Phi_\Omega$ is
\begin{equation}
  \label{Eq16a}
  \Phi_\Omega = \left[ \delta_{i_1} \; \delta_{i_2} \; \cdots \; \delta_{i_M} \right]^T
\end{equation}
Whereas the obtained samples $S_\Omega$ is
\begin{equation}
  \label{Eq16b}
  S_\Omega = \Phi_\Omega \Psi s
\end{equation}
The out-band suppression is conducted by construction of extended projection matrix
\begin{equation}
  \label{Eq17a}
 \bar{\Phi}_\Omega = \begin{bmatrix}
 	\Phi_\Omega \\
 	\left(  \left( \Phi_\Omega^T \right)^C \right)^T
 	\end{bmatrix}
\end{equation}
and extended samples
\begin{equation}
  \label{Eq17b}
 \bar{S}_\Omega = \begin{bmatrix}
 	S_\Omega \\
 	0	\\
 	\cdots \\
	0 	
 	\end{bmatrix}
\end{equation}
Finally, the modified reconstruction expression is given by
\begin{equation}
  \label{Eq18}
   \min \left| s \right|_{l_1} subject \; to \; \; \; \bar{\Phi}_\Omega \Psi s =\bar{S}_\Omega
\end{equation}
The extension of both of the projection matrix and the samples, then put it in the reconstruction expression, will force the out-band components to zero.

\begin{figure}
  \centering
  \includegraphics[width=8cm]{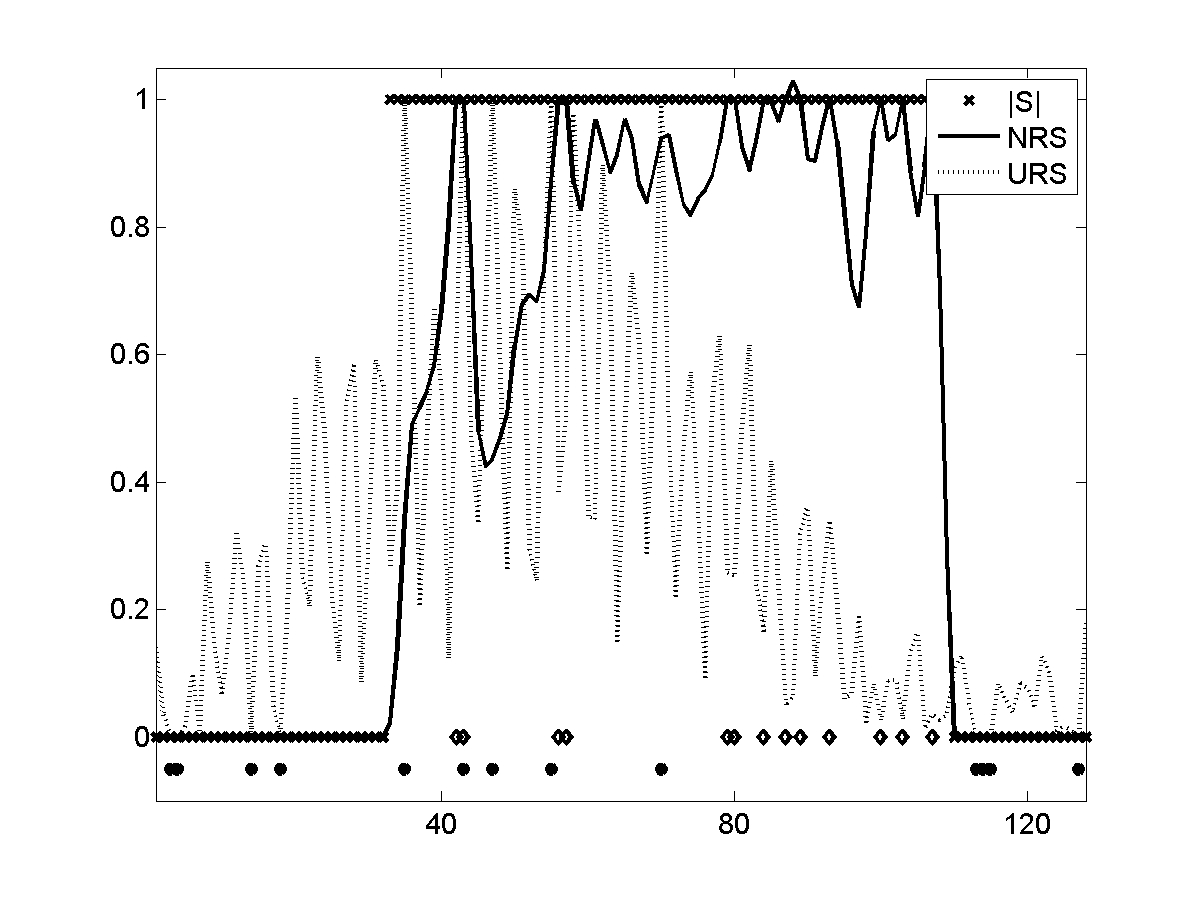}
  \caption{Spectrum of original and reconstructed signal}
  \label{Fig1}
\end{figure}

For clarity, we illustrate this case by the following simulation.  An $N$-length discrete time single-band signal of bandwidth $\left| \Omega \right|=77$, with starting point of the band located randomly (in the instance below, it is 32) is generated. The length of the signal is $N=128$. Two sampling schemes are compared, random sampling without knowing the band limit (URS) and a random sampling within the band. The later case implies taking samples non-uniformly over all of the frequency domain, therefore we call it NRS (Non-uniform Random Sampling).  The spectra of the original and reconstructed signals are displayed in Fig. \ref{Fig1}. The sampling points are shown by asterisk symbols (*) for URS and by diamond symbols for the NRS. We observed that band limitation imposed in the reconstruction (by extension of projection matrix) makes the spectrum reconstructed only within the band, in contrast to non-band limiting case that allows reconstruction outside the band.

\begin{figure}
  \centering
  \includegraphics[width=9cm]{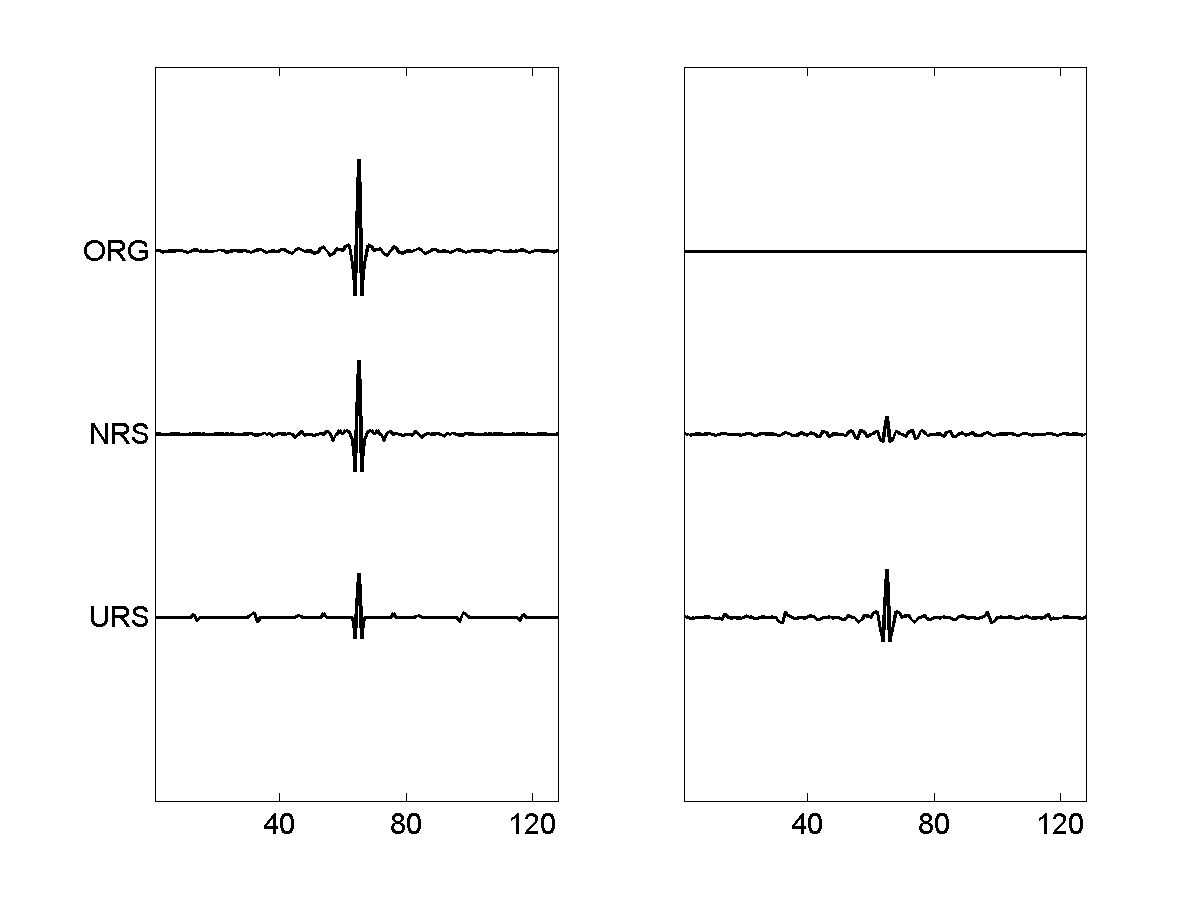}
  (a) \\
  \includegraphics[width=9cm]{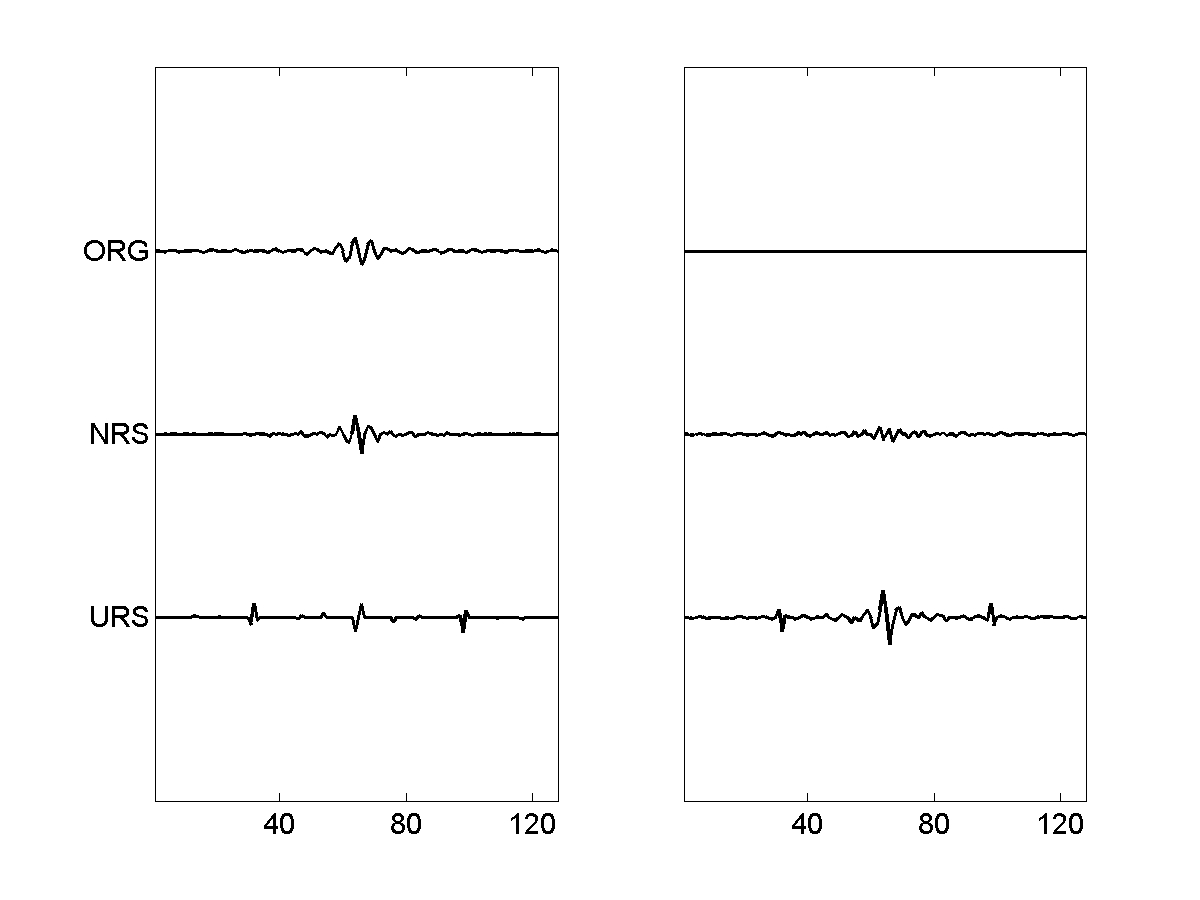}
  (b) \\
  \caption{Reconstructed and error signal. Top: the real part, bottom: the imaginary part, left: original and reconstructed signal, right: residual signal}
  \label{Fig2}
\end{figure}

Signal reconstruction results are shown in Fig.\ref{Fig2}. It is observed that the results for both the real- and imaginary- parts are best reconstructed in the random sampling within the band NRS, which visually also displayed as lower error signal curves in right parts of the figure. Higher error can be attributed to both of less satisfying spectral reconstruction within the band and the overflow outside the band.

%
\subsubsection{Multi-Band Signals}
A multi-band signal is generalization of the single band signal; where, instead of one band, the signal has a number of bands. Without loss of generality, we will consider a two-band signal as an example. Let the first band be $\Omega_1$, while the second one is $\Omega_2$. The number of the coefficients within each bands are $W_1=\left|\Omega_1 \right|$ and $W_2=\left|\Omega_2 \right|$, respectively, and $W_1$ generally differs from $W_2$. As an example, let $W_1 > W_2$. Suppose we take $M$-number of sample from both of these bands in a random-uniform manner and get $M_1$ and $M_2$ respectively i.e., $M_1+M_2=M$. Since $W_1>W_2$, and the sampling is performed uniformly at random, then on the average $M_1>M_2$. Moreover, it will be proportional to the width of the bands or the number of coefficients in the band, therefore, for a repeated $M$-number random sampling in these bands, then
\begin{equation}
  \label{Eq19}
   \left< M_1 \right> : \left< M_2 \right> \approx W_1 : W_2 
\end{equation}
where $ \left< \cdots \right>$ denotes averaging.

We construct the set of delta functions of selected coefficients $D_1$ and $D_2$ from random samples in $\Omega_1$ and $\Omega_2$, respectively, where $|D_1|=M_1$ and $|D_2|=M_2$. Then, the projection matrix $\Phi_{\Omega_1 \cup \Omega_2}$ and its extensions $\bar{\Phi}_{\Omega_1 \cup \Omega_2}$ are
\begin{equation}
  \label{Eq20a}
 \Phi_{\Omega_1 \cup \Omega_2} = \begin{bmatrix}
 	D_1^T \\
 	D_2^T
 	\end{bmatrix}
\end{equation}
and
\begin{equation}
  \label{Eq20b}
 \bar{\Phi}_{\Omega_1 \cup \Omega_2} = \begin{bmatrix}
 	D_1^T  \\
 	D_2^T  \\
 	\left( I-D_1-D_2 \right)^T
 	\end{bmatrix}
\end{equation}
respectively. Whereas the obtained samples and its extension are, respectively,
\begin{equation}
  \label{Eq21a}
   S_{\Omega_1 \cup \Omega_2} =\Phi_{\Omega_1 \cup \Omega_2}  \Psi s
 \end{equation}
\begin{equation}
  \label{Eq21b}
 \bar{S}_{\Omega_1 \cup \Omega_2} = \begin{bmatrix}
 	S_{\Omega_1 \cup \Omega_2}  \\
 	0  		\\
 	\cdots 	\\
 	0
 	\end{bmatrix}
\end{equation}
Then, the $L_1$ reconstruction is done as
\begin{equation}
  \label{Eq21}
   \min \left| s \right|_{l_1} subject \; to \; \; \; \bar{\Phi}_{\Omega_1 \cup \Omega_2} \Psi s =\bar{S}_{\Omega_1 \cup \Omega_2}
\end{equation}

\begin{figure}
  \centering
  \includegraphics[width=9cm]{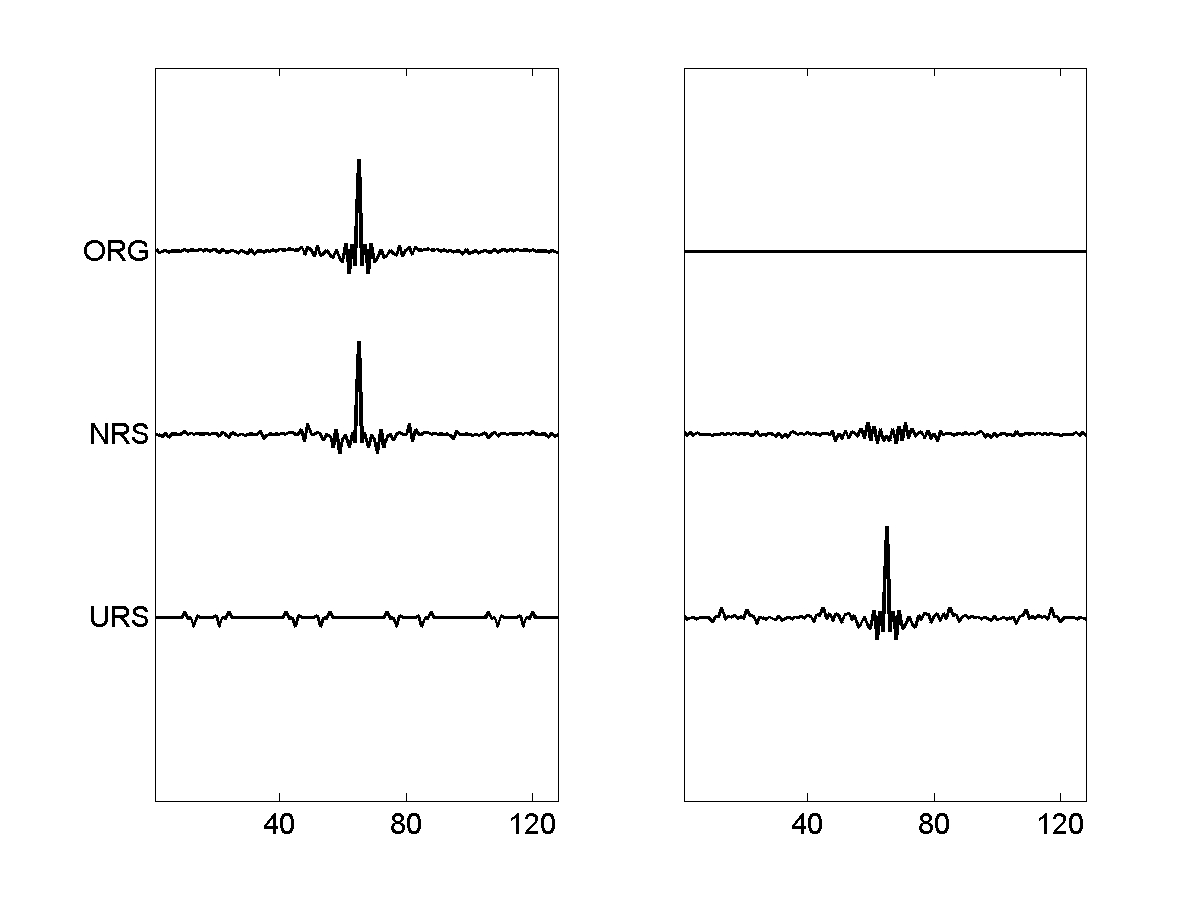}
  (a) \\
  \includegraphics[width=9cm]{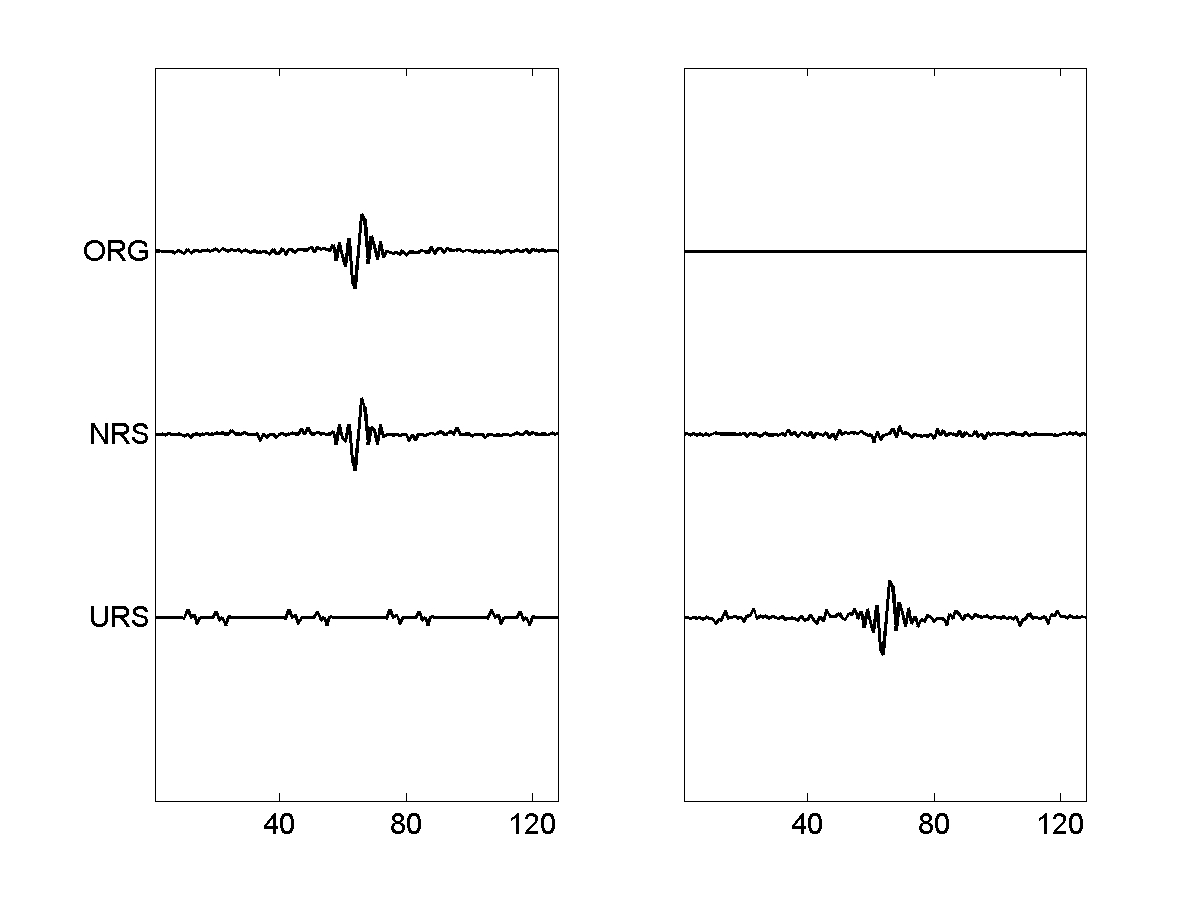}
  (b)
  \caption{Reconstructed and error signals. Top: the real part, bottom: the imaginary part, left: original and reconstructed signal, right: residual signal}
  \label{Fig3}
\end{figure}

\begin{figure}
  \centering
  \includegraphics[width=8cm]{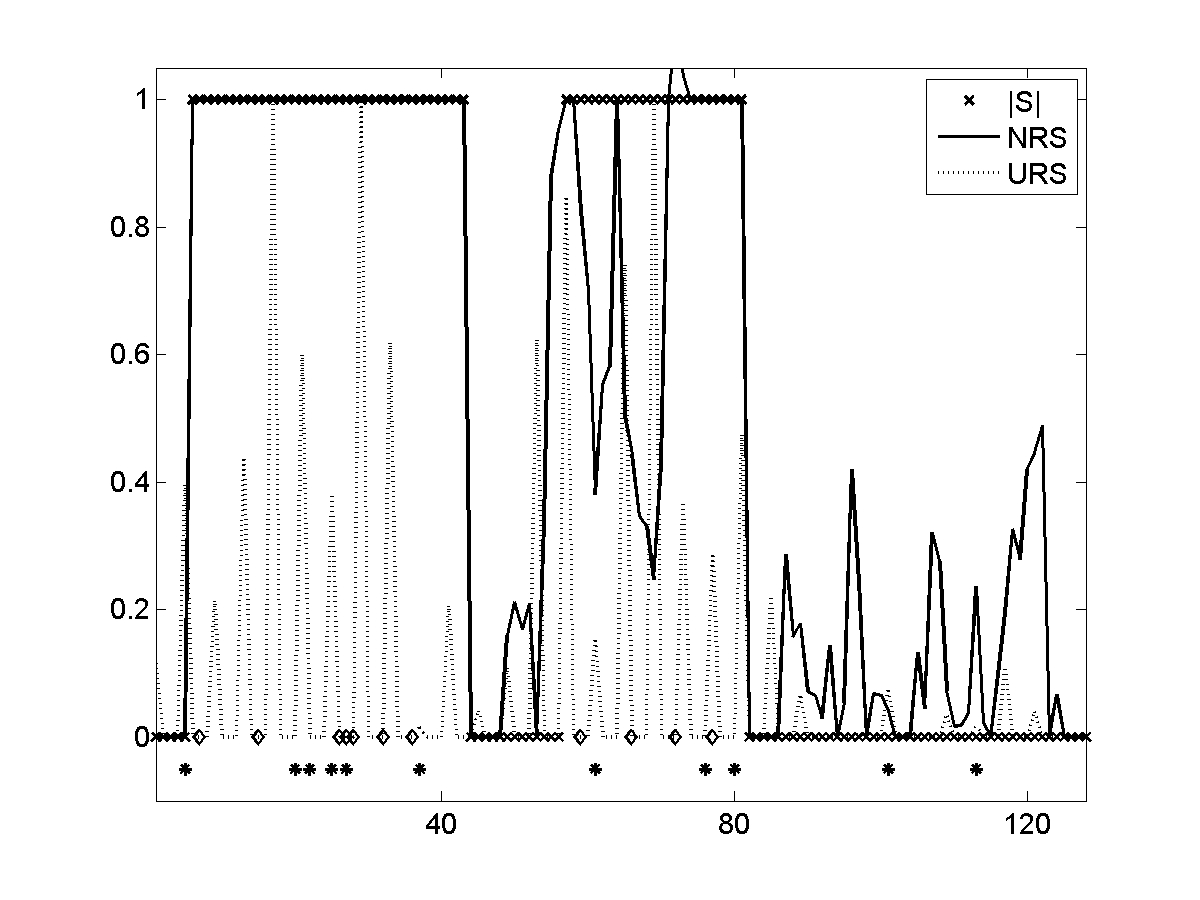}
  \caption{Spectrum of original and reconstructed signal}
  \label{Fig4}
\end{figure}

For illustration, we simulate a two-band signal, where $W_1=38$ and $W_2=25$, then it is uniformly-random sampled with a significantly low sample numbers. Two sampling schemes are compared. The first scheme is by sampling only within these bands with $M_1=7$ and $M_2=4$ denoted as before by NRS. The second one is sampling to the entire bandwidth with $M=11$ number of samples, which is an URS scheme. The results are displayed in Fig.(\ref{Fig3}) and Fig.(\ref{Fig4}).

Fig. \ref{Fig3} shows original and reconstructed signals and corresponding residual error, with (a) the real part and (b) imaginary parts. We observed significant difference for both of these schemes. Whereas the NRS successfully reconstruct the signal and leave small residual error, the URS is not successful due to low number of samples.

The spectra shown in Fig. \ref{Fig4} display the reconstruction performance in frequency domain. The NRS recover most of the spectrum, whereas URS cannot follow the original spectrum. Although spectral leakage happens for both methods, the NRS is relatively smaller than URS.

\subsection{CS of Single- and Multi-Tone Signals}

\subsubsection{Single-Tone Signal}
A single-tone signal $s(n)$ is a signal that is built from a single Fourier coefficient. The spectrum will consisting of a single peak, whose position indicates the location where the measurement should be performed. After obtaining $S(k)$, the signal can easily be  constructed by inverse DFT expressed in (\ref{Eq7b}). Similar stage of CS for band-limited signal can be applied. Since we have only one coefficient and we also know the location, in-band sampling yields only a single coefficient. Assume that it is the $k^{th}$ coefficient. The projection matrix will be consisted of a single row vector $\delta_k^T$ and the projection is done by 
\begin{equation}
  \label{Eq23}
  S_k = \delta_k^T \Psi s
\end{equation}

In the reconstruction process, it is necessary to set all of other frequency components to zero explicitly. Therefore, the projection matrix $\delta_k^T$ should be extended into an $N \times N$ matrix
\begin{equation}
  \label{Eq24a}
  \bar{\delta}_k^T= \left[ \delta_k \; \underbrace{\delta_1 \; \delta_2 \; \cdots \; \delta_{k-1} \; \delta_{k+1} \; \cdots \; \delta_N }_\text{N-1}
  \right]^T \equiv
  \begin{bmatrix}
    \delta_k^T \\
    \left( \delta_k^C \right)^T
  \end{bmatrix}
\end{equation}
Similarly, the observed vector $S_k$ is also extended into 
\begin{equation}
  \label{Eq24b}
  \bar{S}_k= \left[ S_k \; \underbrace{0 \; \cdots \; 0 }_\text{N-1}
  \right]^T 
\end{equation}
Then, the reconstruction of the single tone signal is done by
 \begin{equation}
  \label{Eq25}
   \min \left| \Psi s \right|_{l_1} subject \; to \; \; \; \bar{\delta}_k^T \Psi s =\bar{S}_k
\end{equation}

\subsubsection{Multi-Tone Signal}
This is a generalization of the previous case, where we have $K$-number of non-zero Fourier coefficients. After identifying the locations of $K$-peaks, measurements are conducted at corresponding frequencies. Similar to the single tone case, after getting the Fourier coefficients $S(k)$, the IDFT can directly be applied to recover the signal $s(n)$ exactly. The CS is initialized by constructing a $K \times N$ projection matrix $D^T$ that corresponds to the position of the $K$-peaks,
\begin{equation}
  \label{Eq26a}
  D = \left[ \delta_{i_1} \; \cdots \; \delta_{i_K} \right]
\end{equation}
Accordingly, the extended projection matrix is,
\begin{equation}
  \label{Eq26b}
  \bar{D} = \begin{bmatrix}
    	D^T \\
    	\left( D^C \right)^T
  \end{bmatrix}
\end{equation}
Then, the reconstruction is done by the following optimization
\begin{equation}
  \label{Eq27}
   \min \left| \Psi s \right|_{l_1} subject \; to \; \; \; \bar{D} \Psi s =\bar{S}_D
\end{equation}
where $\bar{S}_D$ is extended $S=D^T \Psi s$ obtained by (\ref{Eq6}).

\subsection{A Brief Summary}
We have seen that URS is consistent with all of the cases. In the single and multiple time-domain cases, sampling is done over all possible Fourier components. Since they are spread over the entire frequency; there is no reasonable preference selecting particular coefficient(s) over the others. In the single and multi-band case, URS is only done on the in-band coefficients, while the reconstruction process is modified by elaborating the knowledge that the out-band coefficients are zero. 

In the single- and multi-tone cases, when the spectrum have peaks at a few places, sampling $M \ll N$ number of samples over $K<M \ll N$ peaks converges into selecting only the $K$-peaks. We can consistently elaborate the exact $K$-number sampling into CS; i.e., by extending the projection matrix and the sampled vector in $L_1$-optimization expression.

Therefore, the URS scheme is consistent to all of these cases. Distribution of magnitude spectrum essentially means distribution of the signal support in the frequency domain. We summarize the results into following two basic principles: 
\begin{itemize}
  \item \textbf{Principle 1}: Sampling should be conducted in a random-uniform manner.
  \item \textbf{Principle 2}: When the support is known, the sampling should only be conducted at the support, while the non-support should be suppressed.
\end{itemize}
These two principles leads to NRS when the distribution of the magnitude is non-uniform, including the multi band cases where the magnitude between the band gaps are zeros, whereas within the bands they are non-zeros.

\section{Non-Uniform Random Sampling and Reconstruction of Non-Sparse Signals}
\label{Section3}

\subsection{Non-Sparse Signals}

\begin{figure}
  \centering
  \includegraphics[width=4cm, height=4cm]{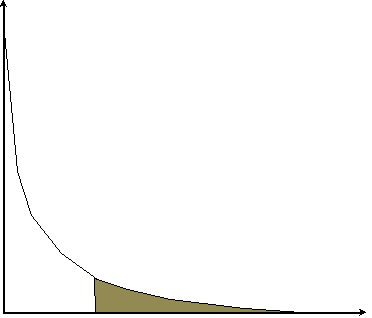}
  \includegraphics[width=4cm, height=4cm]{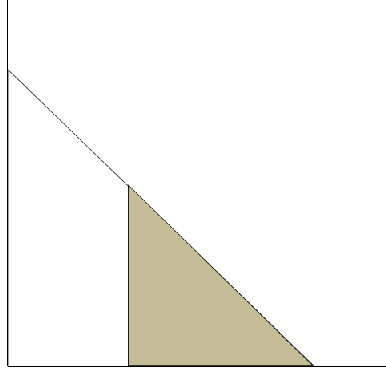}
     $\begin{array}{cccccccccc}
	   (a) & & & & & & & & & (b) \\	
     \end{array} $
  \caption{Typical distribution of sorted-magnitude of non-sparse signals: (a) exponentially decays, (b) linearly decays }
  \label{Fig5}
\end{figure}

The CS assumes that a signal is either sparse or compressible. For the sparse case, CS will give an exact reconstruction result, whereas a compressible signal will be approximated with $K$-sparse signal and leave a small residual error. Compressible signal is defined as a signal whose ordered-magnitude coefficients (in a particular sparsity bases) decays at least exponentially, like one that shown in Fig.\ref{Fig5}(a). Reconstruction error for an approximation with a $K$-sparse signal is shown as a shaded region whose amount decays rapidly with the number of $K$. The case will be different for a non-sparse-non-compressible signals such as a signal whose distribution of ordered-magnitude decays linearly as shown in Fig.\ref{Fig5}(b). After obtaining $K$-largest coefficient, the residual error will be relatively large. 
The similar case is also found for a signal whose magnitude that decays stepped-wisely as shown in Fig.6(a). Assumes that the relative bandwidth for each band $\left| \Omega_i \right|$ are as follows
\begin{equation}
  \label{Eq28}
  \left| \Omega_2 \right| = \left| \Omega_1\right|, \; \left| \Omega_3 \right|= 2 \left| \Omega_1\right|, \; \left| \Omega_4 \right|= 4 \left|\Omega_1\right| \; 
\end{equation}
 Whereas the magnitude of each band is
 \begin{equation}
  \label{Eq29}
  A_k = \left( 2^k-1\right)A
\end{equation}
therefore, $A_4=A, A_3=3A, A_2=7A, A_1=15A$.
  
If the magnitude of the Fourier coefficients is used as consideration in selecting the dominant ones, although the signal is compressible, only the first band will be reconstructed and a lot portion of residual error are remains. 
\begin{figure}
  \centering
  \includegraphics[width=4cm, height=4cm]{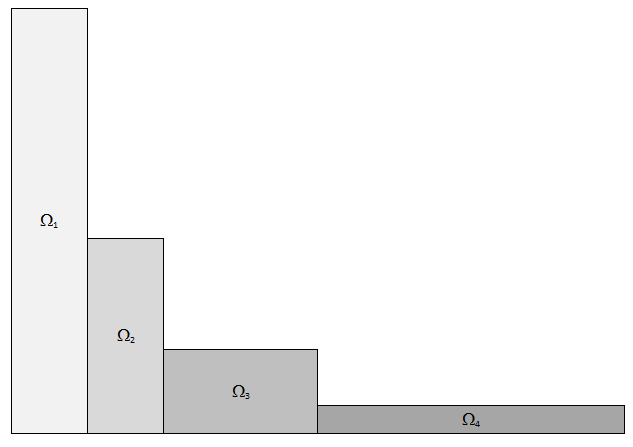}
  \includegraphics[width=4cm, height=4cm]{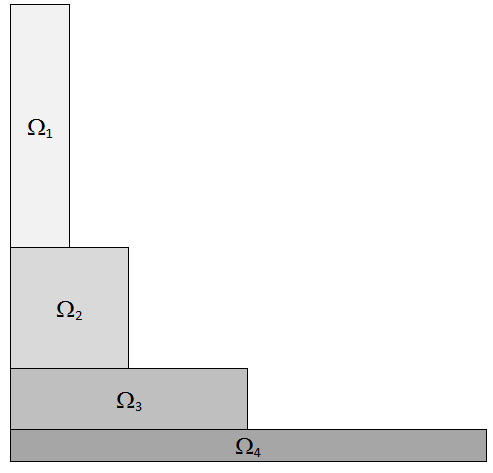}
     $\begin{array}{cccccccccc}
	   (a) & & & & & & & & & (b) \\	
     \end{array} $
  \caption{Two schemes of Spectral partition: (a) into four non-overlapped bandwidth (vertical bands), (b) into four stacks with overlapping bandwidth (horizontal bands) }
  \label{Fig6}
\end{figure}

We can have a better result by considering it as a four-band signal and perform CS for each band, similar to the previous section. Then, the CS for the first band $\Omega_1$ is conducted by uniformly random sampling the band with $M$ number of sampling points followed by $L_1$ reconstruction to this band. Similarly, since the width of $\Omega_2$ equal to $\Omega_1$, i.e., $\left| \Omega_2 \right| =\left| \Omega_1 \right|$ same number of samples $M$ can be allocated. For the third band, since now the bandwidth is doubled, thanks to the UP, the signal $s(n)$ will be sparser than before so that a fewer number of sample than $M$ could be used. We need a signal model to determine the allowed reduce number in the third band more precisely.

Let $S(k)$ be a rectangular function corresponding to band $\Omega_p$ started from $\Omega_{p-1}+1$ and ended at  $\Omega_{p-1}+p$. The magnitude of time-domain signal as given by (\ref{Eq15b}), assuming that $\left| \Omega_p\right|$ is an odd number, will be
\begin{equation}
  \label{Eq30}
 \left| s(n) \right| = \left| 
   \frac{ \sin \left(\pi \left( \left| \Omega_p\right| +1 \right) \frac{n}{N} \right)}
   {\sin \left( \frac{\pi n}{N} \right)} 
 \right|
\end{equation}
The signal width $\Delta t$ can be indicated by the width of the main lobe. Let\rq{}s assume this is the width when the $\left| s(n) \right|$ crosses zero, which happens at
\begin{equation}
  \label{Eq31}
  \frac{\pi}{N} \left( \left| \Omega_p \right|+1 \right) n = \pm \pi \Rightarrow n=\pm \frac{N}{\left| \Omega_p \right| +1}
\end{equation} 
Therefore, the signal width indicating the sparsity of the signal is 
\begin{equation}
  \label{Eq32}
  \Delta t = 2 \frac{N}{\left|\Omega_p \right|+1}
\end{equation}

We see from Eq.(\ref{Eq32}) that increasing the bandwidth $\left| \Omega_p \right|$ by $k$-factor will decrease the sparsity $\Delta t$ to about $\frac{1}{k}$. In turn, it changes the minimum sample number $M$. Therefore, the subsample number for the third band will be $\frac{M}{2}$, whereas for the fourth band, which is four times wider than the first band, will be $\frac{M}{4}$. The total number of allocated sampling points is therefore $M+M+\frac{M}{2}+\frac{M}{4}=2.75 M$.

\subsection{Compressive Sampling Paradox and Non-Uniform Random Sampling}

Summarizing the result of the last discussion, UP helps us to reduce the sample number from $4M$ into $2.75M$. In fact, the UP introduces a sampling paradox: \textit{whereas in classical sampling one should increase the sample number when the bandwidth is increased, the CS suggests the opposite that the number of the sample can be reduced}. Moreover, this paradox suggests a better signal decomposition; instead of vertical decomposition by dividing the signal into band of frequency in Fig.\ref{Fig6}(a), we can horizontally slicing the signal as displayed in Fig.\ref{Fig6}(b). The analysis is as follows.

Again, let’s assume that CS in the first band $\Omega_1$ needs $M$ subsample. The next band, whose bandwidth is twice of the first band, needs $\frac{M}{2}$, the third band needs $\frac{M}{4}$, and the fourth requires $\frac{M}{8}$. This unconventional slicing gives a total of $M+\frac{M}{2}+\frac{M}{4} +\frac{M}{8}=1.875 M$, which is much lower number of subsample than the vertical slicing. Furthermore, since overlapping the bands make it possible to reuse sample in the intersection of the bands. The indicated number in each band is the lower bound, therefore, we may re-use the sample in the previous band for CS in the current band. Now, we need only $M+\frac{M}{4} +\frac{M}{8}+\frac{M}{16}=1.4375 M$. These results are summarized in Table \ref{Table1}, where VD denotes Vertical Division, HD is Horizontal Division, and HU stands for Horizontal division with sample re-use.
\begin{table}
  \caption{Comparison of Sample Allocation For Three Different Schemes}
  \label{Table1}
   \centering
    \begin{tabular}{|c|c|c|c|c|c|}
 	\hline\hline	
	 Method & \multicolumn{5}{c}{Number of Samples in Each Band} \\
	            & Band-1 & Band-2 & Band-3 & Band-4 & Total \\
      \hline
	VD & $M$ & $M$ & $\frac{M}{2}$ & $\frac{M}{4}$ & $2.7500M$ \\
	HD & $M$ & $\frac{M}{2}$ & $\frac{M}{4}$ & $\frac{M}{8}$ & $1.8750M$ \\
	HU & $M$ & $\frac{M}{4}$ & $\frac{M}{8}$ & $\frac{M}{16}$ & $1.4375M$ \\
    \hline  
  \end{tabular}
\end{table}
In the table, we observe a scheme to allocate sample points in signal whose spectrum is step-wisely-decaying as $2^{-i}$, where $i$ denotes the band indices. For this particular case, the sample reuse scheme gives the sample in each band that is also step-wisely decaying.

 \begin{figure}
  \centering
  \includegraphics[width=8cm]{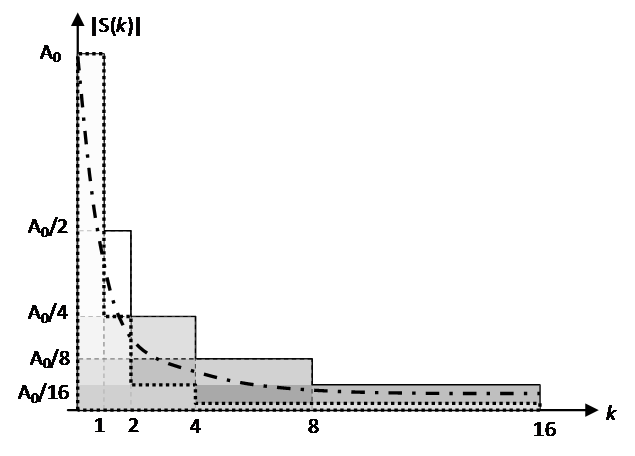}
  \caption{Curves of signal spectra $\left| S(k) \right|=A_0 k^{-\beta}$: step-wise decay with $\beta=1$ shown in solid line, step-wise decay with  $\beta=2$ as dotted line, and any continuous decay with  $1<\beta<2$ shown 
in dashed-dot line }
  \label{Fig7}
\end{figure}

To understand the relationship between the sample density and the magnitude-coefficient distribution, first consider a graph of step-wise spectral decay shown in Fig.\ref{Fig7}.

The values and expression of all related variables are listed in Table \ref{Table2}. We can see that, for a signal with magnitude spectrum decays step-wisely given by
\begin{equation}
  \label{Eq33}
  \left|S\left(k \right) \right|= A_0 k^{-1}
\end{equation}
where $A_0$ is a positive constant indicating maximum value of the magnitude. The density of the sample is
\begin{equation}
  \label{Eq34}
  \rho \left(k \right)= M_0 k^{-1}
\end{equation}
where $M_0$ is the number of allocated sample in the first band. The density is proportional to the square of the magnitude. 
 
Fig.\ref{Fig7} also shows a signal spectrum that decay step-wisely faster than (\ref{Eq33}) as
\begin{equation}
  \label{Eq35}
  \left|S\left(k \right) \right|= A_0 k^{-2}
\end{equation}
The same analysis also gives similar sample density. In fact, for any positive real number $\beta$, the sample density are the same as (\ref{Eq34}).  The difference is when we apply iterative reconstruction, i.e., for a signal with spectrum given by
 \begin{equation}
  \label{Eq36}
  \left|S\left(k \right) \right|= A_0 k^{-\beta}
\end{equation}
The scaling in each of iterations for (dyadic) band division is given by
\begin{equation}
  \label{Eq37}
	E_n=\left( 1-2^{-\beta}\right) E_{n-1}
\end{equation}
\begin{table}
  \caption{Relationship of Variables in Dyadic Band}
  \label{Table2}
   \centering
    \begin{tabular}{|c|c|c|c|c|c|c|}
 	\hline\hline	
	 No & Attribute &\multicolumn{5}{c}{Values} \\
	 \hline
	  1 & $i$ & 1 & 2 & 3 & 4 & 5 \\
        2 & $E_{n-1}/E_n$ & $1/2$ & $1/2$ & $1/2$ & $1/2$ & $1/2$ \\  
        3 & $k$ 			& $1$ & $2$ & $4$ & $8$ & $16$ \\  
        4 & $|S(k)|/A_0$ 	& $1$ & $1/2$ & $1/4$ & $1/8$ & $1/16$ \\  
        5 & $\rho(k)/M_0$ 	& $1$ & $1/4$ & $1/16$ & $1/64$ & $1/256$ \\  
    \hline  
  \end{tabular}
\end{table}
 %
 %

The iterative CS reconstruction is given in an algorithm displayed in Fig.\ref{Fig9a}.
  
\begin{figure}
 \begin{framed}
   \begin{enumerate}
  	\item Preparation
  		\begin{itemize}
  			\item Estimate the total energy $E_0$
			\item Identify $\beta$ of the spectrum profile given by (\ref{Eq36}), set scaling factor $f_\beta$
			\item Set the number of overlapping band $B$
			\item Set the number of sample $M$ and allocation on each bands
			\item Construct projection matrix $\Phi$
			\item Set residual error equal to the signal: $e_{sub}^{curr} = S_{sub}$
			\item Set current estimate to zero:$ \hat{s}=0$;
  		\end{itemize}
	\item Iterative CS: for $m=1$ to $B$
	\begin{itemize}
  			\item $curr_{subsamp}=subsamp(m)$
  			\item Extend both of projection matrix and $curr_{subsamp}$
  			\item Perform $L_1$, obtain solution $s_{sol}$
  			\item Update signal estimate: $\hat{s} = \hat{s} +  f_\beta \times s_{sol}$
  			\item Epdate current error: $e_{subcurr}=e_{subcurr} –(1 - f_\beta) \times \Phi \Psi s_{sol}$
  			\item Update the projection matrix $\Phi$ for next stage
  		\end{itemize}
  \end{enumerate}
\end{framed}
 \caption{Iterative Vertical Slicing Algorithm}
 \label{Fig9a}
\end{figure}
%
\section{Simulations and Analysis}
\label{Section4}
\subsection{Step-wise Power Decay}

\begin{figure}
  \centering
  \includegraphics[width=8cm]{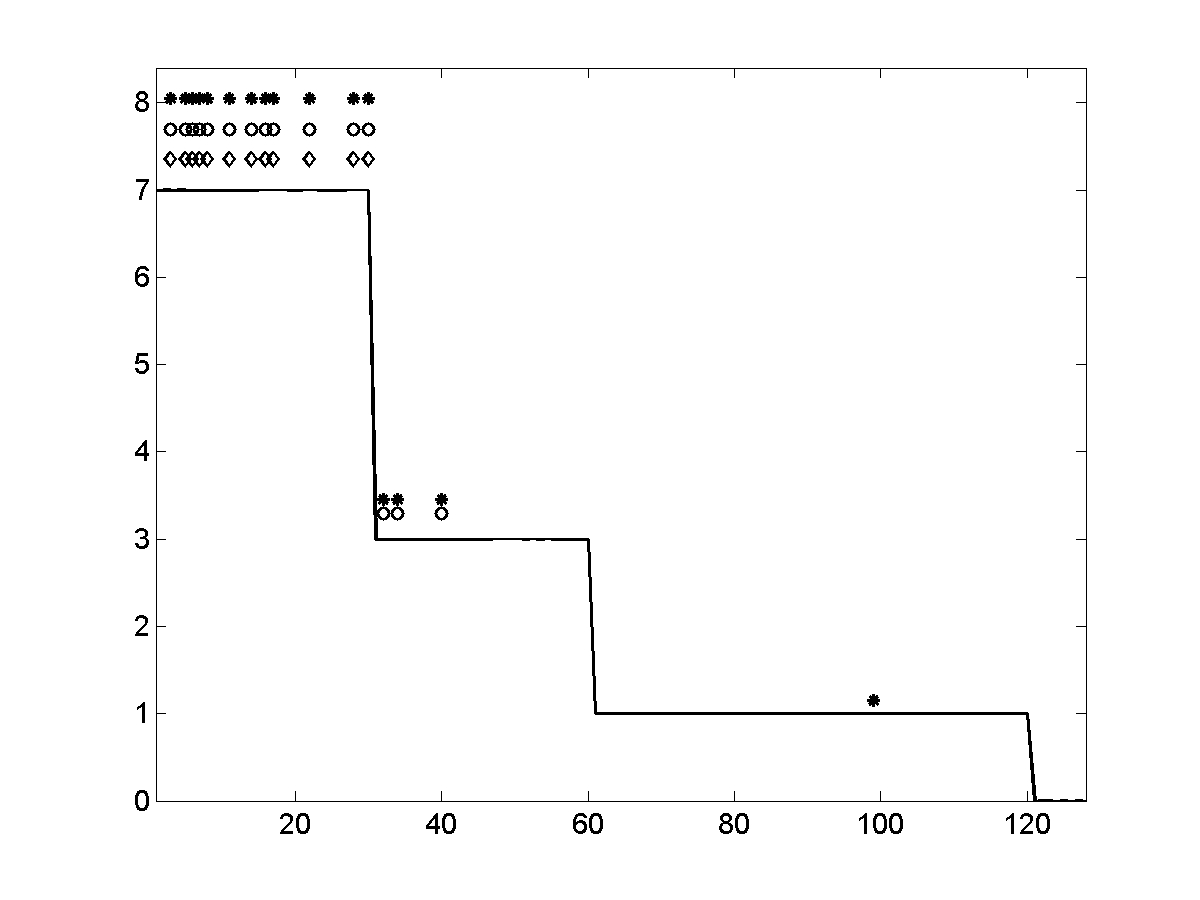}
  \caption{Signal magnitude spectrum and position of the sampling points}
  \label{Fig9}
\end{figure}

In this simulation, we use a three-band signal that has a step-wise power decay spectrum. The bandwidths in vertical sections are 30, 30, and 60 respectively; corresponding to 30, 60, and 120 in horizontal sections. The total number of subsample is restricted to $M=16$, therefore, the horizontal section scheme will allocate the sample vertical-wise for each bands into $\left[ 12 \;\; 3 \;\; 1 \right]$. In the horizontal slicing reconstruction with reuse sample, this allocation corresponds to $\left[ 12 \;\; 15 \;\;16 \right]$ subsamples. An example of sampling points configuration is shown in Fig.\ref{Fig9}. 

The signal is reconstructed iteratively, in this case into three iterations. In the first iteration, 12 number of subsample shown as diamonds in Fig.\ref{Fig9} is used. Since in this special case all of the horizontal bands have identical energy, only $1/3$ of the reconstructed signal is retrieved in the first and second iterations. Similarly, $15$ samples denoted by circle are used in the second reconstruction, and all of the $16$ samples are used in the final iteration.
 \begin{figure}
  \centering
  \includegraphics[width=9cm]{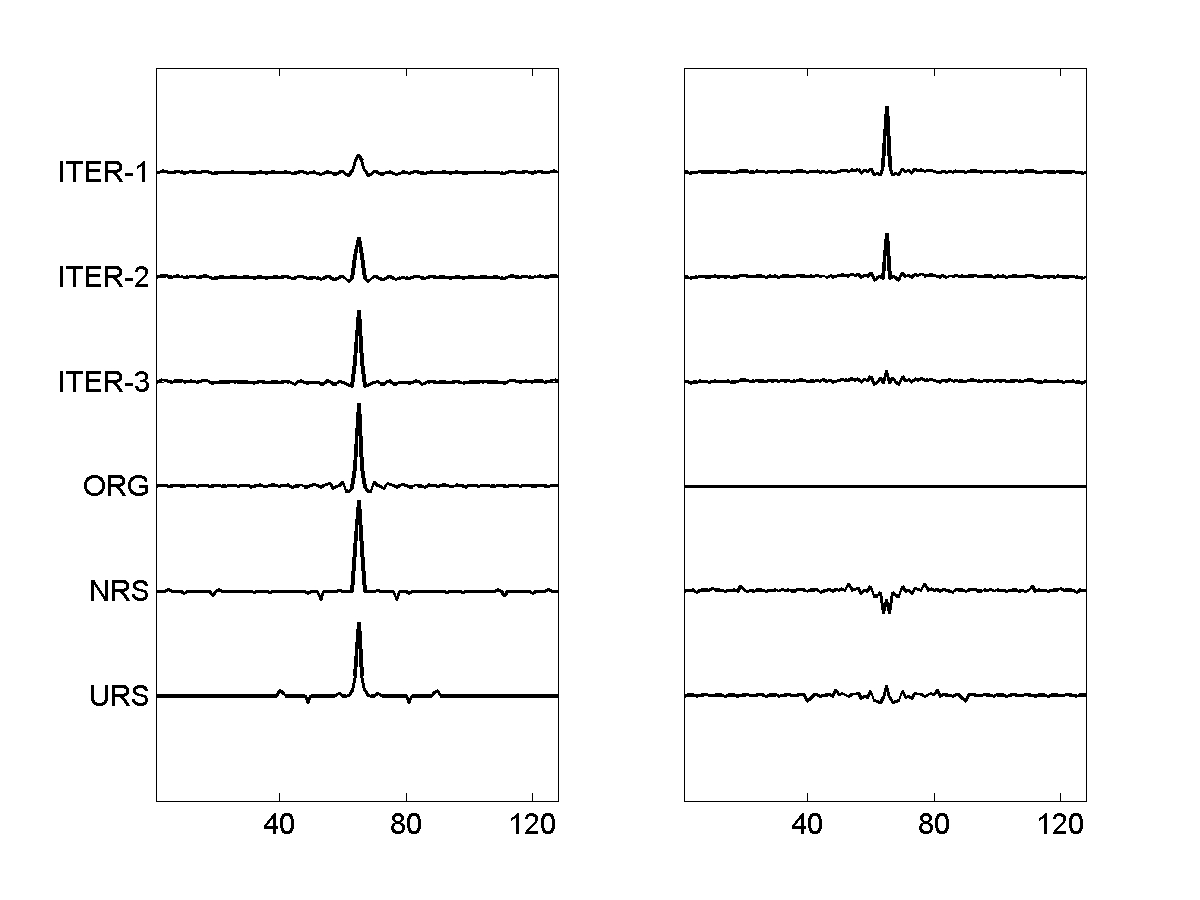}
  (a) \\
  \includegraphics[width=9cm]{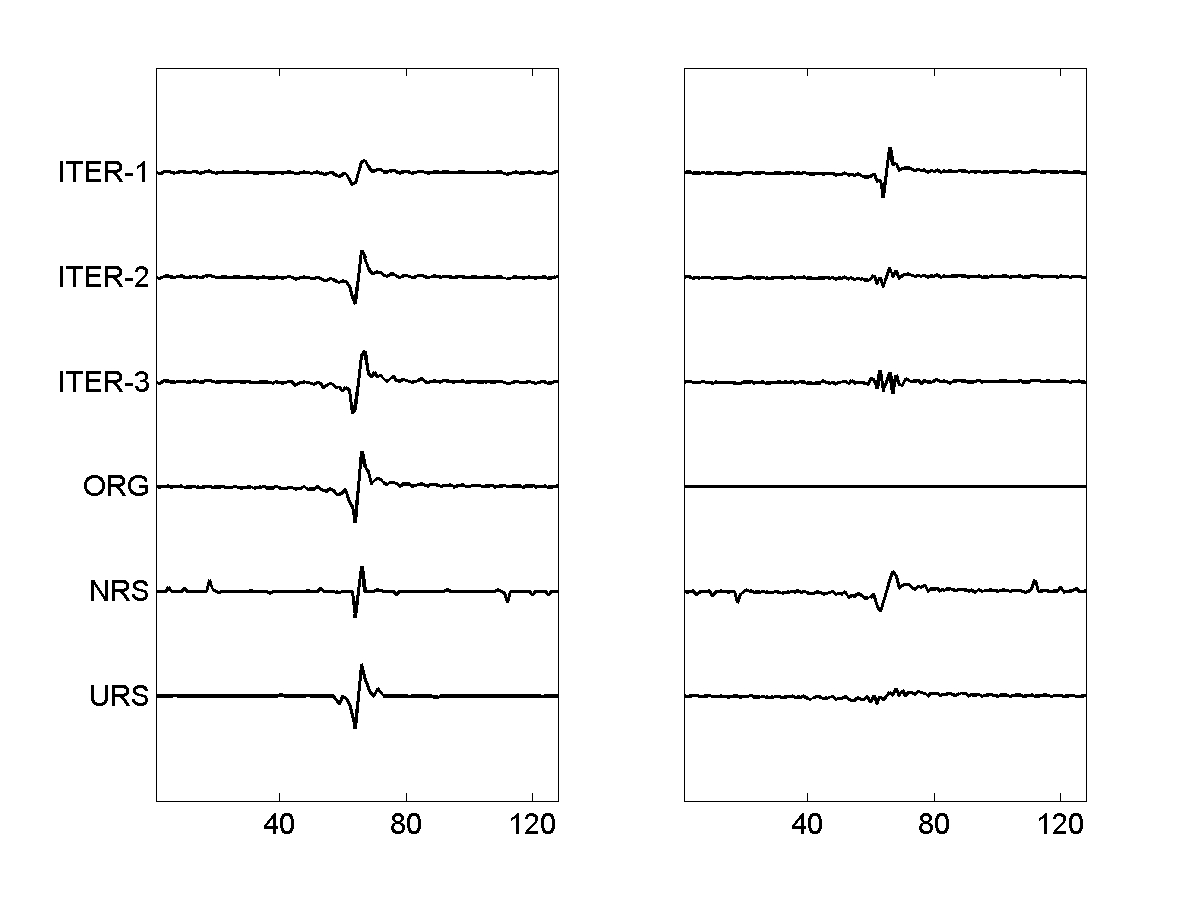}
  (b) \\
   \caption{Reconstructed and residual error signal: (a) real-part, (b) imaginary part}
  \label{Fig10}
\end{figure}

Fig.\ref{Fig10} shows the original signal and reconstruction results by three methods. Original signal is labeled ORG, the usual random-uniform sampling with URS, three stages in iterative non-uniform sampling indicated by ITER-1, ITER-2, and ITER-3, and non iterative non-uniform samping by NRS.  For both of the real and imaginary part, the reconstruction result becomes better with the iteration, which is also indicated by decreasing magnitude of residual error signal. After $10$ times simulation, the average RMSE (Root Mean Square Error) of each methods are, $RMSE_{URS}=0.206$, $RMSE_{NRS}= 0.167$, and $RMSE_{RNW}= 0.094$. 
 \begin{figure}
  \centering
  \includegraphics[width=8cm]{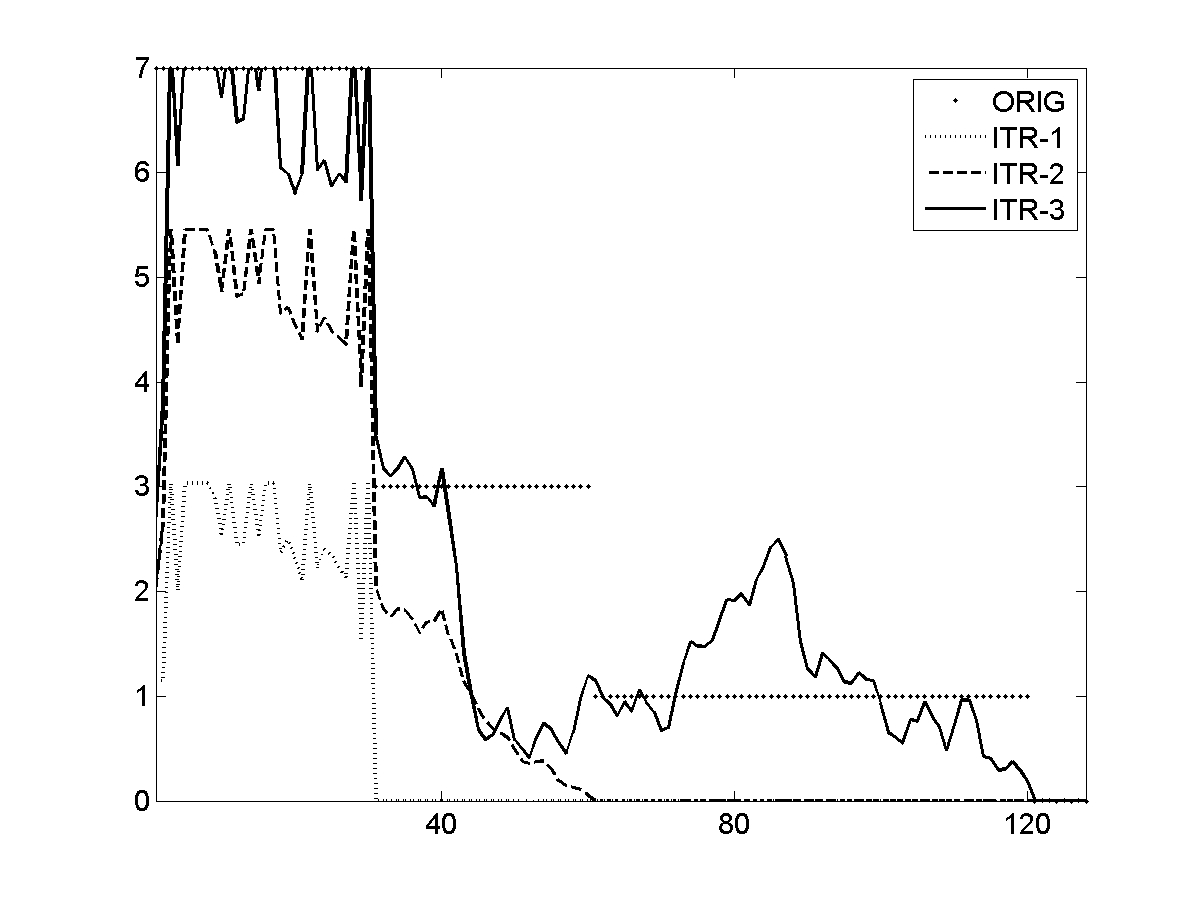}
  \\ (a) \\
  \includegraphics[width=8cm]{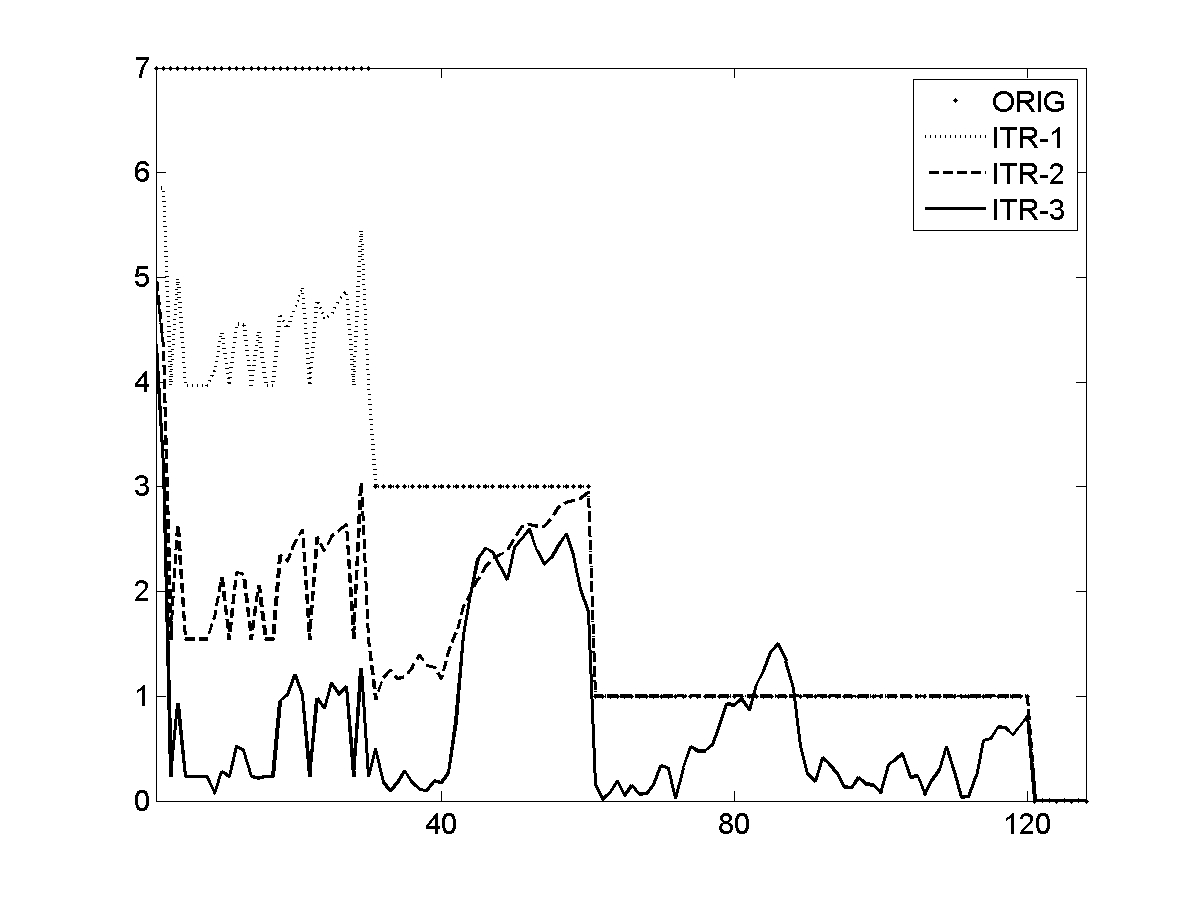}
  \\ (b) \\
  \caption{Original and reconstructed spectrum for each iteration: (a) reconstructed spectrum, (b) spectrum of residual error}
  \label{Fig11}
\end{figure}

We can see that iterative method give the best results, while the NRS is still better than URS. It shows that, even without systematic horizontal slicing, allocating the sample non-uniformly proportional to the energy density, i.e. sum of magnitude coefficient divide by the bandwidth, is better than just performing random uniform sampling.

The improvement is also shown in the frequency domain in Fig.\ref{Fig11}. The magnitude spectrum becomes closer to the original signal spectrum, whereas the error spectrum in the right part of the figure reduces in each of the iteration.

\subsection{Step-Wise Linear Decay}
\begin{figure}
  \centering
  \includegraphics[width=8cm]{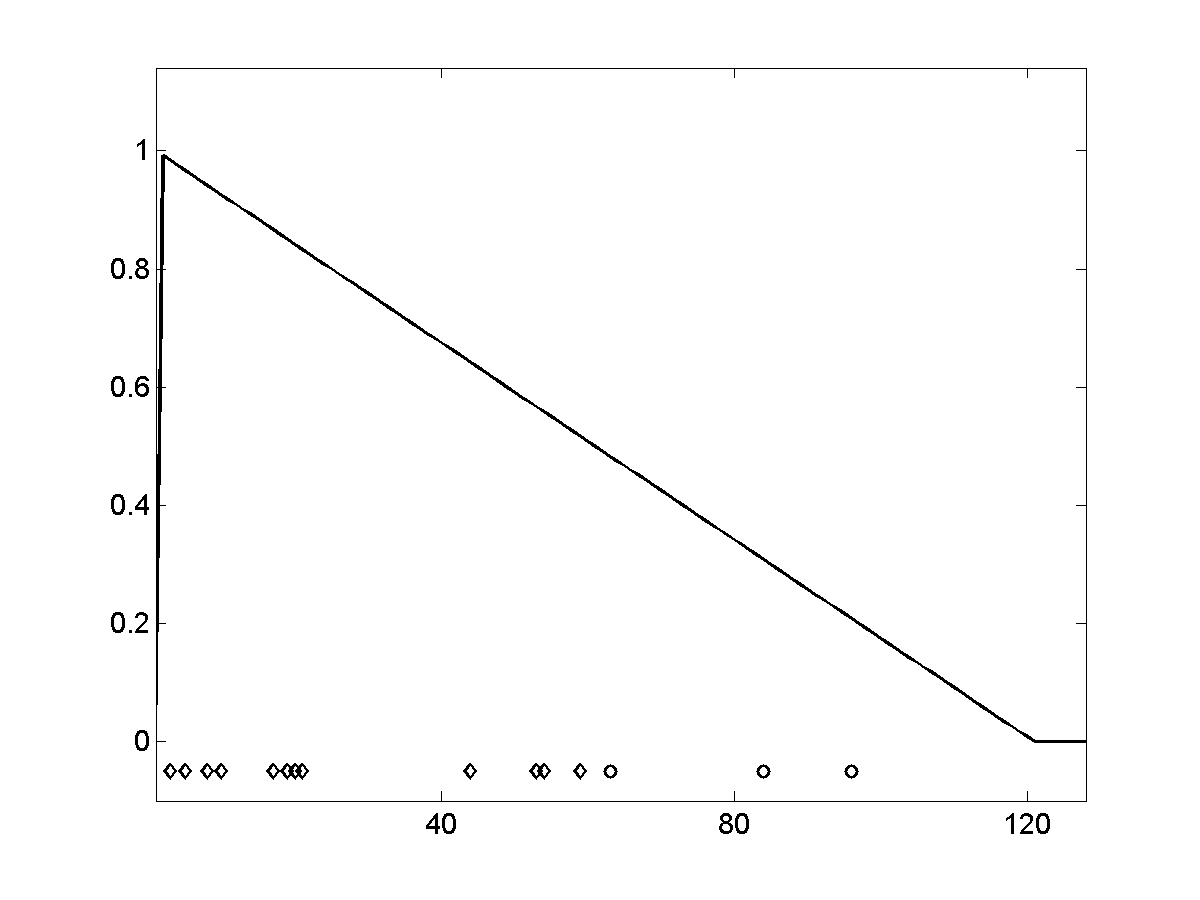}
  \caption{Signal magnitude spectrum and position of the sampling points}
  \label{Fig12}
\end{figure}

In this simulation, we perform iterative CS for the signal that has triangle-shape spectrum. The total bandwidth is also 120 which are divided into two bands. The number of sampling points is 15, where 12 points are allocated for the first band and 3 points are for the second one. Sampling point allocation scheme is displayed in Fig.\ref{Fig12}.

\begin{figure}
  \centering
  \includegraphics[width=9cm]{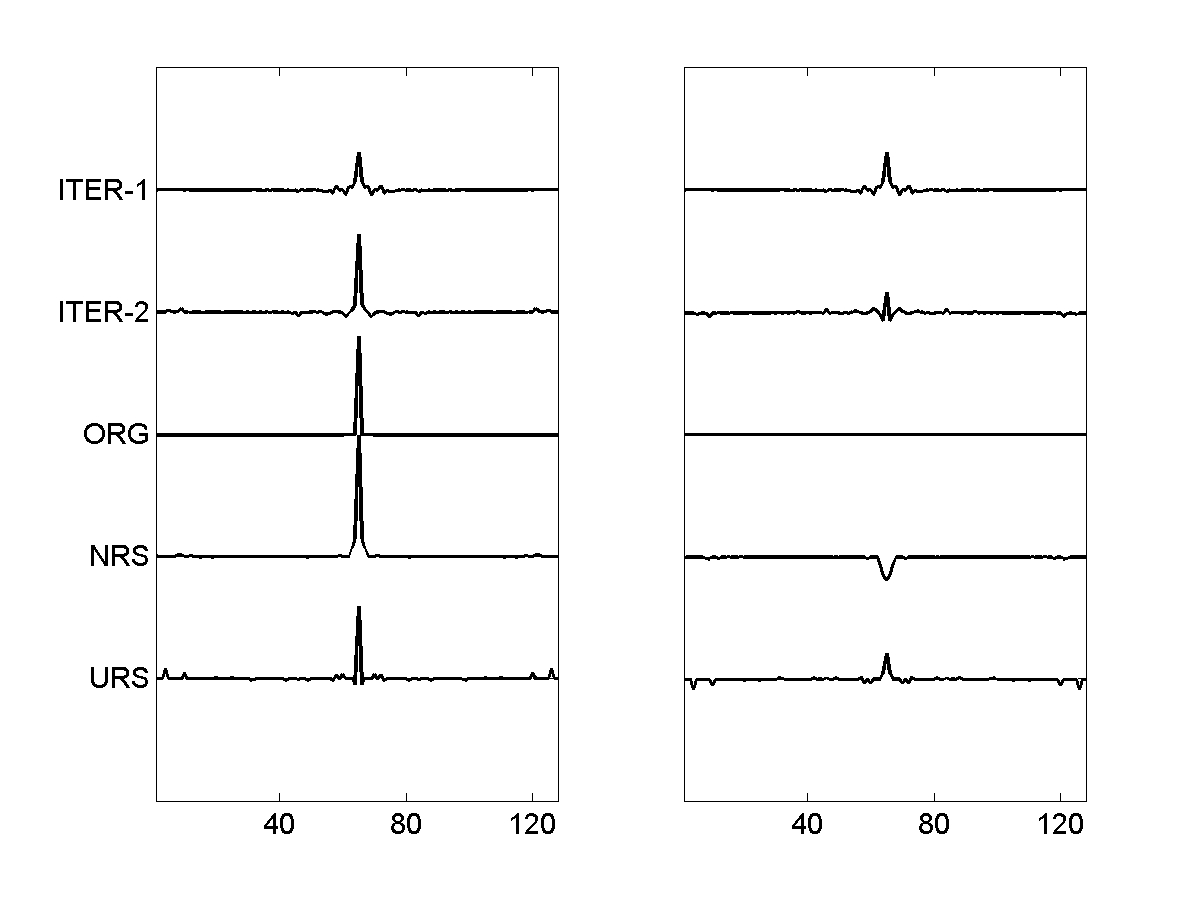}
  (a) \\
  \includegraphics[width=9cm]{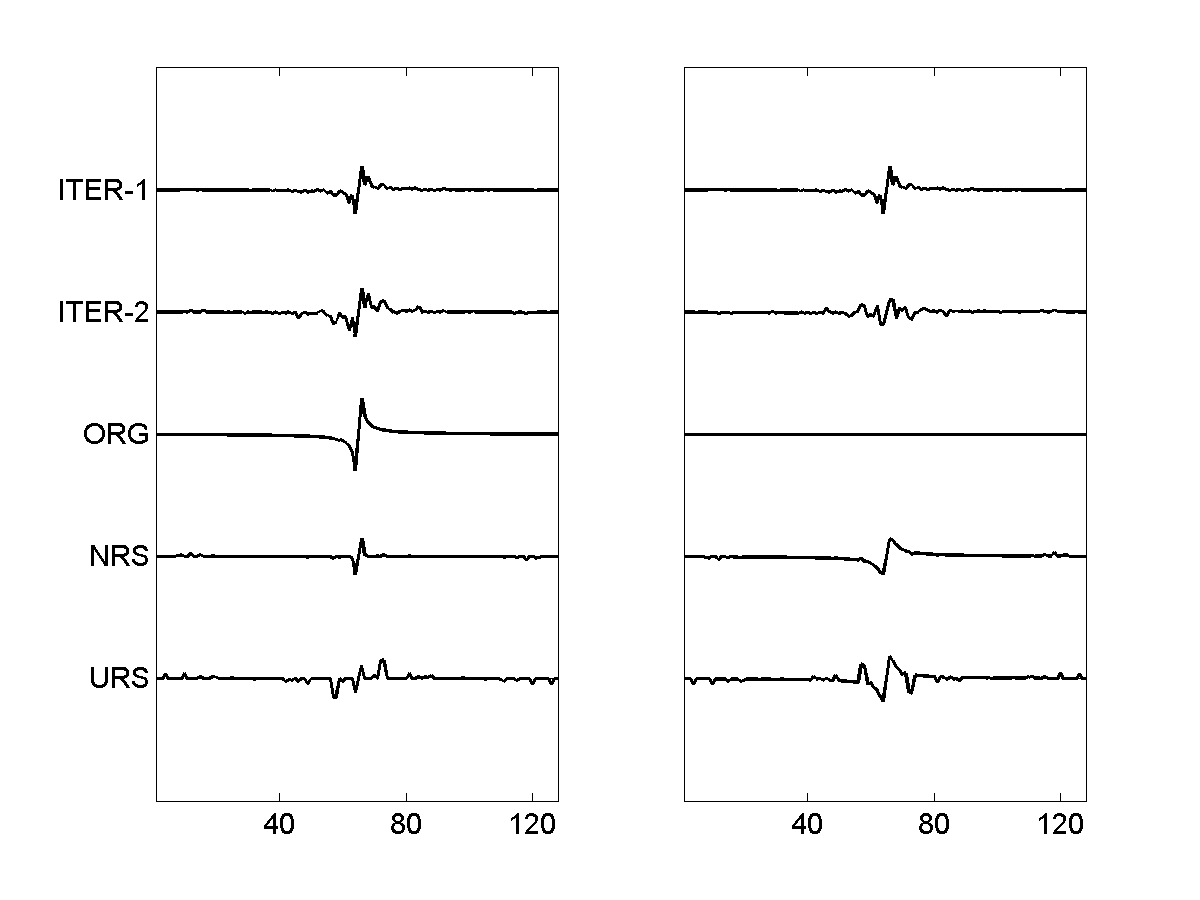}
  (b) \\
  \caption{Reconstructed and Residual error signal: (a) real-part, (b) imaginary part}
  \label{Fig13}
\end{figure}

Iterative horizontal slicing is now performed. In the first stage, 12 number of sample is used. Then additional 3 samples are used to CS reconstruct the signal with 15 samples. The result shown in Fig.\ref{Fig13} displays an improvement using this method. Similarly, the result is compared with URS and NRS. We repeat the reconstruction 10 times and obtain the following RMSE values of each method, i.e., $RMSE_{URS}=0.0234$, $RMSE_{NRS}=0.0201$, and $RMSE_{RNW}=0.017$. Again, the iterative methods perform best and the non-iterative non-uniform sampling performs better than URS.

\begin{figure}
  \centering
  \includegraphics[width=8cm]{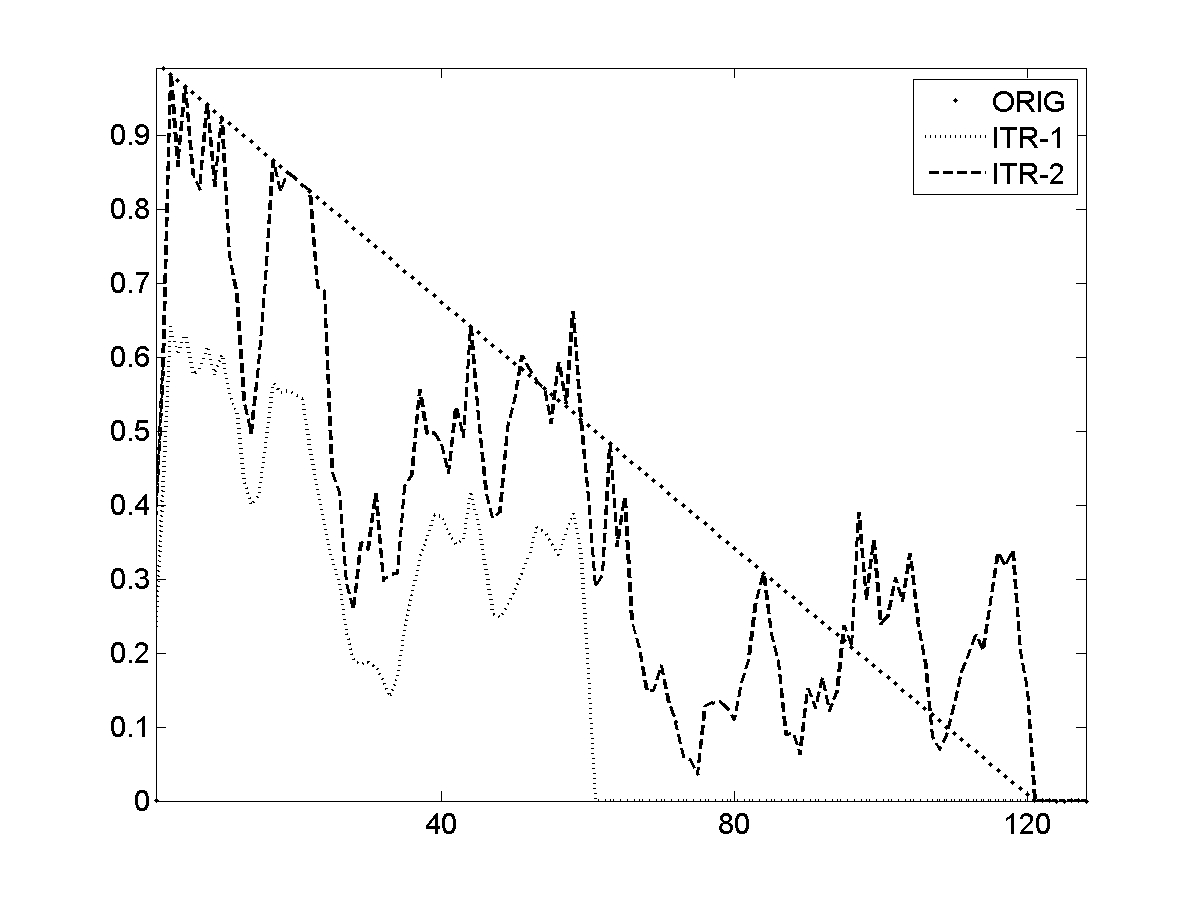}
  \\ (a) \\
  \includegraphics[width=8cm]{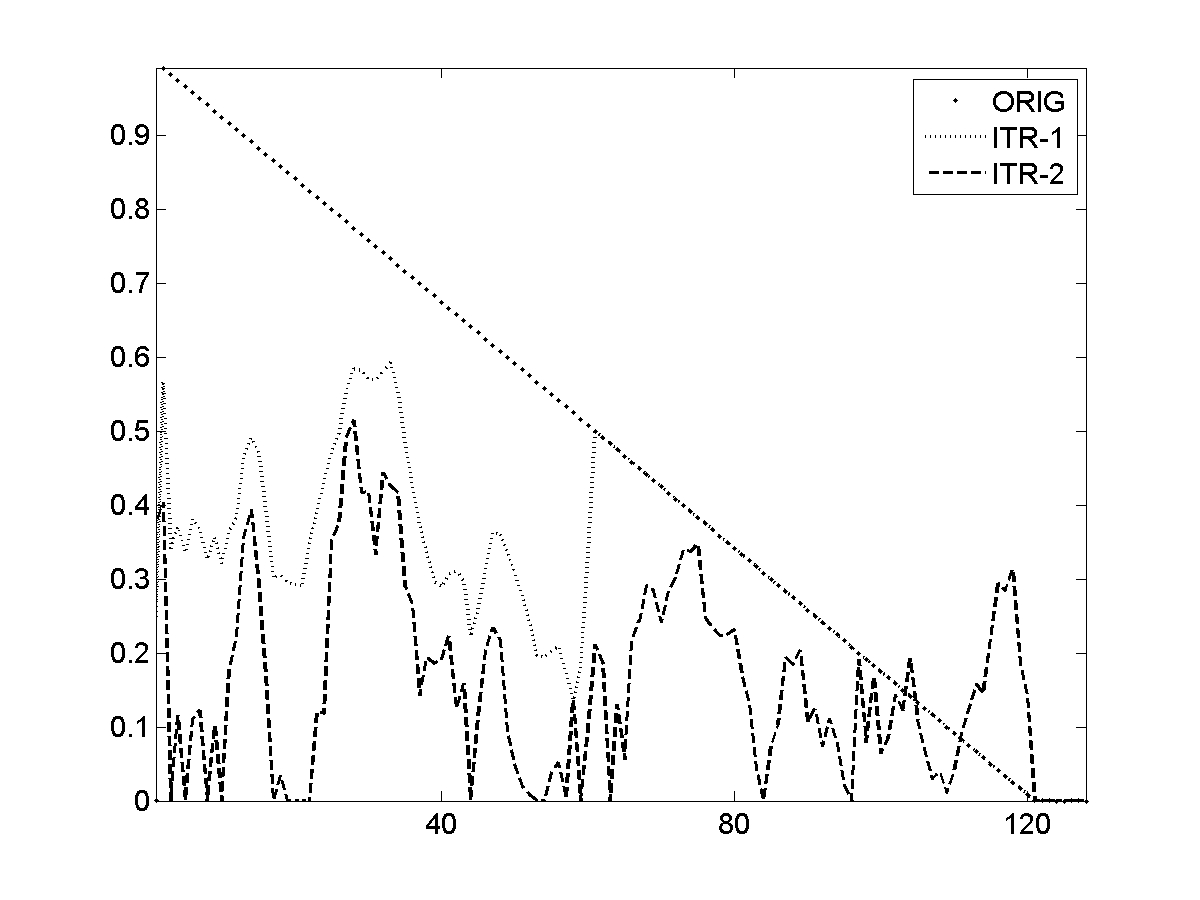}
  \\ (b) \\
  \caption{Original and reconstructed spectrum for each iteration: (a) reconstructed spectrum, (b) spectrum of residual error}
  \label{Fig14}
\end{figure}

Further analysis in the frequency domain, displayed in Fig.\ref{Fig14}, shows how the iterative method reconstructs the spectrum. In the last stage, most of spectrum is retrieved and the error spectrum is reduced.

\section{Conclusion}
\label{Section5}
We have presented an analysis on sample allocation problem in compressive sampling. We obtain an important principle extended from uniform-random sampling (URS) that the random sampling should be conducted within the signal support and non-supports should be suppressed in the reconstruction process b explicitly in optimization program. Further, URS and the UP for extended spectrum case leads to non-uniform sampling, in which high magnitude coefficients should be allocated more samples than low magnitude ones. This result is compatible with conventional sampling, such as in transform coding.


\ifCLASSOPTIONcaptionsoff
  \newpage
\fi


\end{document}